\documentclass[12pt,letterpaper]{article}
\usepackage{amsmath,amssymb,pgf,pgfarrows,pgfnodes,float,appendix, hyperref}
\usepackage{graphicx}
\usepackage{subfigure}
\usepackage[margin=0.9in]{geometry}

\title{{\rm\footnotesize \qquad \qquad \qquad \qquad \qquad \ \qquad \qquad \qquad \ \ \ \ \ \                  RUNHETC-2020-34, UTTG-07-2020}\vskip.5in Primordial Black Holes as Dark Matter}
\author{Tom Banks\\
Department of Physics and NHETC\\
Rutgers University, Piscataway, NJ 08854\\
E-mail: \href{mailto:tibanks@ucsc.edu}{tibanks@ucsc.edu}\\ 
W. Fischler\\
Department of Physics \\
University of Texas, Austin,TX, 78712\\
E-mail: \href{mailto:fischler@utexas.edu}{fischler@utexas.edu}}

\date{July 30,2020}
\begin{document}
\maketitle

\begin{abstract}  We investigate models in which a spectrum of black holes with Hawking temperature of order the radiation temperature at the beginning of the radiation dominated era can survive long enough to produce a matter dominated era at the observed crossover between matter and radiation in our universe. We find that a sufficiently dense population of such black holes can indeed do so.   The stronger observational constraint, that the black holes have lifetimes at least as long as the current age of the universe is harder to assess, because of black hole mergers during the matter dominated era.  We then investigate whether the required densities and masses are consistent with the Holographic Space-time (HST) model of inflation.  We find that they are, but put mild constraints on the slow roll parameter $\epsilon = - \frac{\dot{H}}{H^2}$ in that model to be small.  The bound is no stronger than the observational bound on the model's prediction for tensor fluctuations.  The required black hole density, at the reheat temperature, in a model with a single species of black hole, must be viewed as a quantum mechanical accident.  In such a model, our universe exists because of a low probability quantum fluctuation.

\end{abstract}

\section{Introduction}

There has been a lot of renewed interest in the possibility that dark matter consists of primordial black holes\cite{pbh}.  In this paper we will investigate a rather minimal version of that hypothesis.  Almost all modern cosmological models have a period of inflation, after which the universe is reheated and we have a radiation dominated era.
We will assume that at the reheating temperature, we also have a spectrum of black holes with masses $m_{iR}$ and number densities $n_{iR}$.   The number densities are small enough that, especially when Silk damping is taken into account, we can neglect black hole collisions and mergers.   The equations governing the evolution of the universe are
\begin{equation} dm_i / dt =\frac{g}{2} \frac{1}{15360\pi}  (e - \sum_i m_i n_i) m_i^2 - \frac{1}{15360\pi} m_i^{-2} , \end{equation}
\begin{equation} dn_i / dt = - 3n_i\sqrt{{\frac{8\pi}{3}}e} , \end{equation}
\begin{equation} de/dt = - 3 \sqrt{{\frac{8\pi}{3}}e} (e + \frac{1}{3} (e - \sum_i m_i n_i)) .  \end{equation}  $e$ is the total energy density of the universe.  $g$ is the effective number of relativistic degrees of freedom in the radiation.

During the period when the matter energies are negligible, these equations simplify to 
\begin{equation} e dm_i / de =\frac{g}{61040}\sqrt\frac{3}{2{\pi}^3 }( - e^{1/2} m_i^2  + e^{-1/2} m_i^{-2}) , \end{equation}
\begin{equation} n_i = n_{iR} (e/e_R)^{3/4} . \end{equation}  These equations are written in Planck units, where $G_N = 1$. The first term in the mass evolution equation is accretion of radiation onto the black hole, while the second represents Hawking radiation.  In these equations if $m_i^*$ is the value of the mass when the Hawking term begins to dominate, then the mass decays to zero explosively at a time of order $(m_i^*)^3 $, after which the equation is not trustworthy.  Note also that, in Planck units, $e_R \ll 1$ in reasonable cosmological models.  Thus although their are initial values of $m_{iR}$ and $e_R$ for which one can prove that the Hawking rate never dominates, they are not of cosmological interest.  

The question of whether sufficient accretion can occur to substantially increase the black hole mass was answered in the negative by Novikov and Zeldovich\cite{nz}.  To see this, drop the Hawking term, and write the solution
\begin{equation} m_i = \frac{m_{iR}}{1 - \frac{g}{30720}\sqrt\frac{3}{2{\pi}^3 } m_{iR} (e_R^{1/2} - e^{1/2})}  . \end{equation}   As long as $m_{iR} e_R^{1/2} \ll 1$ there is no appreciable increase of the mass with decreasing energy density.  This means that we can make an assumption that simplifies the solution of the full set of equations.  This is the assumption that the black hole decay time scales like $m_{iR}^3$ and the black holes are all far from each other.   Each black hole then decays as if it were in flat space, and adds entropy to the universe which, depending on $n_{iR}$ might raise the temperature back above the radiation temperature at the decay time.
Taking this possible reheating into account we can then see which black holes survive long enough to dominate the energy density.  There's an interesting question of whether this repeated reheating would leave detectable signals in the CMB.

The simplest case, on which we concentrate in this note, has just a single black hole mass at the reheat temperature. This case applies whenever all PBHs lighter than the largest one decay much earlier, without significant reheating of the universe. In this case we find a lower bound on the number density $n_R$, such that matter radiation equality occurs before the black holes decay.  Requiring that the energy density at which this occurs is the observed $(1 eV)^4 = 10^{-112}$ gives a relation between $e_R, n_R$ and $m_R$.    

The idea of primordial black hole dark matter is a natural one in the HST model of inflation because the immediate post inflationary era is dominated by a dilute gas of black holes with Schwarzschild radius $m$ of order the inflationary Hubble radius.  The radiation dominated era follows from the decay of most of these black holes.  However, long before this occurs, density fluctuations in the black hole gas go non-linear, which is likely to lead to the creation of larger black holes through merger.   These would be the "primordial" black holes of the title.   We will call the black holes of mass $m$ {\it inflationary black holes } (IBHs). It is difficult to calculate the spectrum of PBH masses without doing extensive N-body simulations of IBH mergers before decay.  

The HST model calculates $e_R$ in terms of the inflationary black hole mass $m$, and the slow roll parameter $\epsilon$ of the inflationary era.  When we impose the constraint of getting the right density at matter radiation equality, we get another determination of these parameters, which is consistent with the determination from the CMB if the slow roll parameter is not too large.   In the HST model, the tensor to scalar power spectrum ratio is proportional to $\epsilon^2$, but the numerical coefficient is not determined and might actually depend on the details of the underlying quantum gravity model.  {\it Thus, it is plausible that the HST model of inflation can consistently fit all of the requirements of cosmological history through the era of matter radiation equality}. The problem that remains to be solved is to determine whether mergers of PBHs after matter radiation equality can eliminate a population of small black holes whose Hawking decay is ruled out by observation\cite{pbh}.

\section{Single species model}

Let the black hole mass be $m_R = a e_R^{-1/4}$ with $a > 1$, but not too large.   Its decay time is thus
\begin{equation} t_d = \frac{10240}{g}\pi a^3 e_R^{-3/4} . \end{equation}  Since $m_R e_R^{1/2} \ll 1$ this is a very long time.   The radiation density after some long time $t$ is
\begin{equation} e = e_R t^{-2} ,\end{equation} while the black hole energy density is
\begin{equation} e_b = m_R n_R t^{-3/2} . \end{equation}  These are equal at
\begin{equation} t_{eq} = (an_R)^{-2} e_R^{5/2} .\end{equation}
The energy density at matter radiation equality is
\begin{equation} e_R^{-4} (a n_R)^4 . \end{equation}   The observed density of equality implies that 
\begin{equation} a n_R =  10^{-28} e_R. \end{equation} 
This calculation is valid if 
\begin{equation} t_{eq} < t_d , \end{equation}  so that the black holes have not yet decayed at $t_{eq}$. Thus we have the inequality
\begin{equation} \frac{10240}{g}\pi a^3 e_R^{-3/4} >  t_{eq} = (an_R)^{-2} e_R^{5/2} . \end{equation}
\begin{equation} a^3 > \frac{g}{10240\pi}10^{56} e_R^{5/4} . \end{equation}

Black holes of mass $a e_R^{-1/4}$ will be stable on times of order the current age of the universe if  $e_R \sim (\frac{10240}{g}\pi)^{4/3}10^{-80} a^4 \sim (\frac{2}{g})^{4/3} (2.5\ {\rm GeV}\ a )^4 $.  This is not a reasonable value, but during the period of black hole domination density fluctuations grow and black holes can merge and become more stable.  Determining the value of $e_R$ that would produce primordial black holes consistent with observational bounds  requires numerical simulations to understand the merger history of PBHs after PBH matter domination begins.

Up until this point, our considerations have been independent of any particular model for how PBHs and the CMB radiation originate.  We now turn to a particular model.

\subsection{PBH Dark matter in the HST model}

We will review the HST model of inflation in the penultimate section of this paper.  For now, it is enough to recall that in this model, the immediate post-inflationary state of the universe, on lines of constant FRW time, is a dilute gas of black holes of average mass $m$ with a number density of order $m^{-2}$, just below the threshold where the  holes would coalesce and merge.  The individual black holes are modeled as large quantum systems in a Hilbert space of dimension $e^{4\pi m_+^2}$ with a fast scrambling\cite{hpss} Hamiltonian whose matrix elements between typical states are of order $1/m$.   The systems are thus in thermal equilibrium.  $m_+$ is a bit larger than $m$ and there are small fluctuations around the average mass, $m$, of relative order $\frac{\delta m}{m} \sim 1/m$.
These fluctuations appear on the past horizon of a post-inflationary timelike geodesic as scalar curvature fluctuations.  There are similar fluctuations of angular momentum around its mean of zero, which account for the tensor fluctuations on the sky.  These are of the same order of magnitude, as one can estimate from the entropy formula for Kerr black holes.  The angular momentum fluctuations show up as a rotating black hole and gravitational radiation carrying off the compensating angular momentum.  By the usual rules of statistical mechanics, the fluctuations are approximately Gaussian, with $k-$point functions behaving like $m^{-k}$ .   In co-moving gauge, which we've implicitly been using, the usual gauge invariant measure of scalar fluctuations is $\zeta = \epsilon^{-1} \delta R$, with $\epsilon \equiv - \frac{\dot{H}}{H^2}$.   Thus the scalar tensor ratio $r$ is
$$ r \propto \epsilon^2 .$$
In single field inflation models, scalar and tensor fluctuations come from quantum fluctuations of the metric in co-moving gauge, calculated using the Einstein Lagrangian.   This reduces the enhancement of the scalar fluctuation power spectrum by a factor of $\epsilon$, compared to the HST formula, and allows us to calculate the numerical coefficient in $r$ in those models.  Unfortunately we have no similar calculation of the numerical coefficient in the HST model.  In addition, we have so far found that the restrictions of unitarity, causality and "relativity" (see the penultimate section) do not put strong constraints on the detailed form of the Hamiltonian, so the coefficient in $r$ might be model dependent.  It is like the ratio between two different susceptibilities in a condensed matter model. We should note however that the conventional definition of $r$ contains a kinematic factor of order $10$ from summing over angular momentum modes of the CMB.  In field theory inflation models $r$ is proportional to $\epsilon$ with a coefficient that is this kinematic factor times a number of order $1$.  Thus, we should probably expect a similar kinematic enhancement of $r$ by a factor of ten compared to the naive ratio of power spectra in HST models.  
Matching the size of the scalar power spectrum to CMB data, we find
\begin{equation} 10^{-5} = \frac{1}{m\epsilon} . \end{equation} 

As noted above, we call the black holes of mass $m$ {\it inflationary black holes} (IBHs).  The IBHs decay at a time $m^3$ and reheat the universe to (see below) 
\begin{equation} e_R \sim m^{-8} . \end{equation}  Just before reheating, the number density of IBHs was $n_{IBHR} \sim m^{-9} $.   However, the dilute black hole gas is a pressureless system, so fluctuations grow and go non-linear at a time $t_{nl}$ given by
\begin{equation} 1 = 10^{- 5} t_{nl}^{2/3} = (m\epsilon)^{-1} t_{nl}^{2/3} . \end{equation}
\begin{equation} t_{nl} = (m\epsilon)^{3/2} , \end{equation} long before the black holes decay.  We remind the reader that the early IBH matter dominated era is not part of the subsequent radiation dominated one and hence there is no radiation damping of the fluctuation growth during that era . Thus there is plenty of time for these black holes to coalesce and form larger black holes, which are the primordial black holes (PBHs) of the previous subsection.  

Without dedicated N-body simulations, it is hard to estimate the number density $n(m_R)$ of black holes formed by these mergers, but we can impose the phenomenological constraints of the previous subsection with relative ease.  The correct energy at matter radiation equality is obtained if
\begin{equation} a n(a e_R^{-1/4}) = e_R 10^{-28} . \end{equation}
The ratio of PBHs to IBHs just before reheating is thus
\begin{equation} \frac{n(a e_R^{-1/4} )}{n_{IBH}} \sim a^{-1} m 10^{-28} = (a\epsilon)^{-1} 10^{-23} .\end{equation} 
Neglecting black hole velocities, a PBH must be formed by an agglomeration of 
$ a e_R^{-1/4}/m  = a m$ IBHs.   The previous equation tells us that only a very small fraction of all collections of $a m$ IBHs need to form a PBH in order to account for the observed crossover between matter and radiation.

The constraint on these models coming from insisting that the PBHs actually remain stable down to $t_{eq}$  is
\begin{equation}   a^3 >  \frac{g}{10240\pi}10^{56} e_R^{5/4} =  \frac{g}{10240\pi}10^{56} m^{-10} \sim10^2 \epsilon^{10}  . \end{equation} In the last equation we've assumed that the fraction involving $g$ is order $1$.  This is probably an overestimate, and it is more likely that this fraction is of order $10^{-1} - 10^{-2}$.
If we use this overestimate and assume only the kinematic enhancement of $r$ over the power spectrum ratio in the HST model, then current CMB data bounds $\epsilon < 10^{-1/2}$ .   A discovery of tensor modes just below this level would turn the above bound into
\begin{equation} a > 0.1 , \end{equation} which is not at all stringent.  Since we take a cube root in writing the bound for $a$, the additional uncertainty about the value of $g$ does not change this conclusion.   In the next section we will review all of the features of the HST inflation model and argue that they are consistent with a unified description of inflation, reheating, baryogenesis and dark matter, in terms of black hole physics.

 \section{Review of the HST model of inflation}
 
 Although the HST inflation model is based on a complete non-singular quantum mechanical model of cosmology, most of its important properties are coarse grained 
 features of the quantum model, and can be understood using a few familiar semi-classical ideas.  We will sketch the quantum mechanics in an appendix, since it is needed to argue for approximate $SO(1,4)$ invariance of the fluctuations we see in the CMB.  There are three unfamiliar semi-classical notions that one must accept in order to appreciate the model
 \begin{itemize}
 \item The Covariant Entropy principle (CEP) which states that the maximal entropy acessible in a causal diamond is equal to one quarter of the maximal area of a leaf in a null foliation of the boundary of the diamond (the area of the holographic screen), in Planck units.  In particular, to describe a universe that is future asymptotically de Sitter, we only need a finite dimensional Hilbert space.
 \item To describe cosmology as seen by a detector following a timelike trajectory, one should use a slicing of space-time into space-like slices that fit between nested causal diamonds along that trajectory, all of which start at the beginning of time.  The future tips of consecutive diamonds are separated by a planck time. Each trajectory has its own Hilbert space and evolution operator.  The Hamiltonians are time dependent, because of this causal choice of slicing. The quantum version of the principle of relativity is that the maximal area causal diamond in the intersection of two diamonds must be identified as a subsystem (tensor factor) of each Hilbert space, and the density matrices assigned to that subsystem by each trajectory's dynamics must have the same entanglement spectra.   It must be emphasized that these time slices are NOT the FRW slices of an FRW spacetime.  This principle will be used to argue that the correct post inflationary universe is an FRW model filled with a dilute gas of black holes of average Schwarzschild radius equal to the inflationary horizon size.
 \item Any localized object in the bulk of a causal diamond of entropy $\sim N^2$ is a constrained state in a Hilbert space whose generic state is thought of as living on the horizon and thus, due to redshift, has a Hamiltonian of order $1/N$.  The number of constrained q-bits is $EN$ where $E$ is an approximately conserved quantum number when $N$ is large.  In constrained subspaces the Hamiltonian has additional terms of order $E$, which describe the localized objects when they pass close to the timelike geodesic connecting the tips of the diamond.  The motivation for this principle comes from the Schwarzschild dS entropy formula and the large change in entropy that occurs when a localized object is dropped into a black hole.  Both suggest that a localized object separated from a horizon lives in a Hilbert space of dimension exponentially larger than the tensor product of the separated objects, with the degrees of freedom that mediate interaction between them frozen by a constraint.
 \end{itemize}
 
 The three points above assumed a particular background space-time.  We view that space-time, following Jacobson\cite{ted} as a hydrodynamic description of our quantum system, with statistical quantum fluctuations (Brownian motion) of order the inverse square root of the entropy of de Sitter space, and quantum interference effects 
exponential in that entropy.  Larger, localized fluctuations, will be treated explicitly, and are responsible for the CMB.   This space-time has a family of time-like geodesics and we impose the FRW property on our {\it model} by insisting that the dynamics along each of these geodesics is quantum mechanically identical.  This puts no restriction at all on the initial state.

A model where the initial state is generic can be shown to lead to a flat FRW model with a scale factor 
\begin{equation} a(t) = \sinh^{1/3} (3t/R).  \end{equation}    In each causal diamond, at all times, the entropy is maximized, so that, by the third item above, there are never any localized excitations.  The de Sitter entropy formula for eternal de Sitter space shows that for a given entropy deficit, the maximal entropy is a single black hole.  However, in our cosmology, on the time slicings of item 2, this is not the maximal entropy configuration.  
\begin{figure}[h!]
\begin{center}
  \includegraphics[width=16cm]{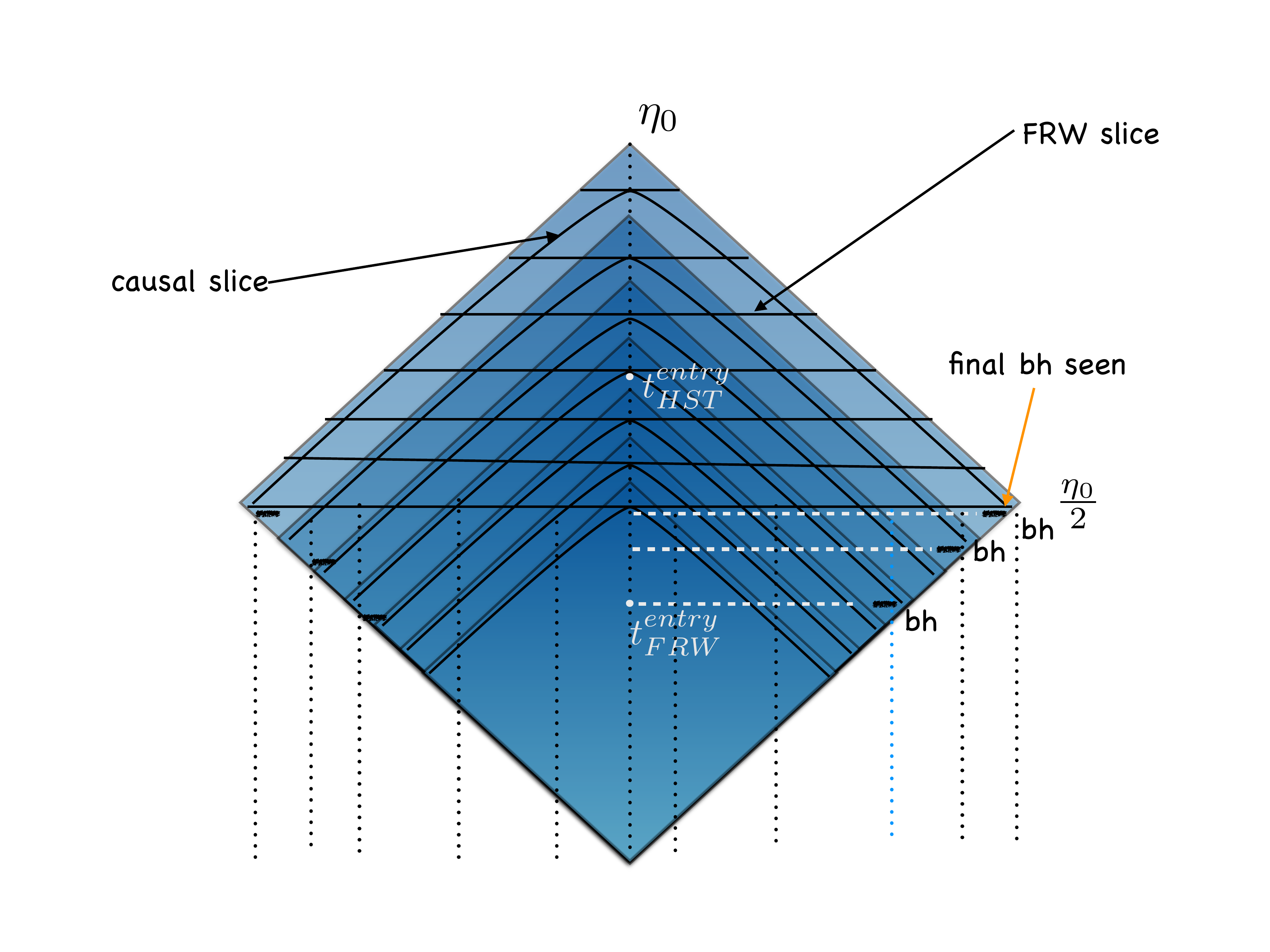}
\end{center}
\vspace{-1.6cm}
\caption{FRW and causal slicings\label{fig1}}
\end{figure}

Figure 1 shows the slicings along a particular geodesic, and the way in which they intersect other geodesics.  Geodesics enter the past boundary of a given causal diamond on different slices.  A black hole of mass $M$ that becomes visible to the chosen trajectory on the red slice is an isolated quantum system with a Hilbert space of dimension $e^{\pi M^2}$.  It is a fast scrambling system\cite{hpss} in a typical equilibrium state.  Consistency of this description with the quantum relativity principle of item 2 implies that the evolution along the blue geodesic must have taken place in the same size Hilbert space, with the same density matrix eigenvalues.  According to item 3 above, this means that the system could equally well be thought of as the horizon of dS space, which is the same as saying that the blue trajectory underwent a period of inflation.  Our choice of a homogeneous model means that the same must be true along {\it every} geodesic, including the one that defines the sequence of diamonds in the figure.

In keeping with the Jacobsonian view of space-time as a hydrodynamic average of a high entropy quantum system, the black hole/dS area is an average entropy, $S$, and all average quantities should have fluctuations of order $S^{-1/2}$.   The Kerr black hole entropy formula tells us that for fixed mass, $L = 0$ is the peak of the angular momentum distribution, and there are Gaussian fluctuations of precisely this order around the mean.  These must be interpreted as a stochastic background of gravitational waves around each black hole\footnote{In models with extra massless particles, there could also be a stochastic background of those particles.}.  The fact that the fluctuations are approximately Gaussian, with $k-$point correlations of order $S^{-k/2}$ follows from general theorems in statistical mechanics.   Detailed questions like numerical ratios between different types of fluctuations are model dependent.  In hydrodynamic treatments of condensed matter systems, ratios like this are encoded in phenomenological "susceptibilities", which are usually difficult to calculate from microscopic first principles.

The phrase "each black hole" in the previous paragraph refers to the fact that the quantum principle of relativity implies that there is a separate black hole along each one of a sufficiently dilute set of geodesics in the FRW space-time, which is equivalent to saying that every geodesic underwent the same period of inflation, with the same Hubble radius up to statistical fluctuations.  To be more precise, we should think of Figure 1 as the causal diamond of an FRW space-time in conformal coordinates.  The period before the midpoint is the inflationary era {\it including the slow roll period where the horizon expanded to its current coordinate position}.   The immediate post-inflationary slice $\eta = \eta_0 / 2$, is a universe filled with a dilute black hole gas, where dilute means that the black holes are sufficiently well separated that they do not suffer significant mergers until the density fluctuations grow to order $1$.  This means in particular that they can be thought of as isolated quantum systems even though some of them are in causal contact in the lower half of the figure.

Our fast scrambling assumption allows us to rephrase this constraint as a bound on the slow roll parameter.  Thermalization in a fast scrambling system with typical time scale $ H^{-1}$\footnote{This time scale can be read off the behavior of the quasi-normal modes of black holes with negative specific heat and/or the dS horizon.  It is incorporated in an explicit factor in the Hamiltonian in the microscopic models described in the appendix.} is completed in a time $- H^{-1} {\rm ln }\ H$.   During slow roll $H(t)$ changes, although $2m$, the inflationary horizon size/ average black hole radius, does not.  The maximal entropy of the system must grow sufficiently rapidly that individual black holes do not have time to thermalize with each other.  That is
\begin{equation} \epsilon > \frac{1}{{\rm ln}\ (mH^{-1} (t))} \end{equation}

The post inflationary universe is thus a flat\footnote{Flatness in HST is again the choice of a model.  The flat FRW model with fixed equation of state has a conformal Killing vector and our HST model of the very early universe, with $p=\rho$ incorporates that symmetry.  It turns out to be important to understanding the approximate $SO(1,4)$ symmetry of the CMB.} $p=0$ FRW model, composed of a dilute gas of black holes of average mass $m$.  The two point functions are approximately $SO(1,4)$ invariant (see the appendix) except that the scalar power spectrum contains a factor of $\epsilon^{-2} (t_k)$, which depends on $k$ through the horizon crossing time of a particular fluctuation.  The scalar spectrum is therefore red-tilted, and the function $\epsilon (t)$ on which we have very few theoretical constraints can be used to fit the details of the spectrum.  Although the functional form is different from that of field theoretic inflation models, we can get an identical fit, because both formalisms have an otherwise unconstrained function of time.    

The scalar tensor ratio $r$ is proportional to $\epsilon^2$, so modulo the uncertainty due to possible large numerical coefficients in the HST models, a given observational bound on $r$ leads to weaker upper bounds on $\epsilon$.  The overall magnitude of the power spectrum leads to the equation 
\begin{equation}10^{-5} = (m\epsilon)^{-1} ,  \end{equation} though this equation also has (possibly model dependent) coefficients.

The IBHs decay at a time $  \frac{10240}{g}\pi  m^3$ after the end of inflation, but density fluctuations in the IBH gas become non-linear in a time $t_{nl} = 10^{15/2} = (m\epsilon)^{3/2} $.  This is much shorter than $t_{d}$, so black hole mergers will occur, producing a spectrum of PBHs at the reheat time $t_R$.  The PBH of the previous sections will be the largest of these.  It is important to calculate the full spectrum of PBHs, as well as the density $n_R$ at $t_R$, in order to get a detailed picture of this cosmology and determine whether there are any novel observational signals.  This requires dedicated N-body simulations, incorporating some estimate of black hole merger rates, which 
we have not done.

Black hole decay can also lead to baryogenesis\cite{tbwfbaryo}.  In that reference we argued that the "naive dimensional analysis" estimate of the baryon number to entropy ratio produced in IBH decay was
\begin{equation} \frac{\Delta b}{\sigma} = \epsilon_{CP} g^{1/4} m^{-3/2} , \end{equation} where $\epsilon_{CP} $ is the effective CP violation in black hole decay and $g$ the effective number of degrees of freedom with mass below the reheat temperature.  The reheat temperature is 
\begin{equation}T_R = g^{-1/4} e_R^{1/4} = g^{-1/4} (\frac{d_I }{ m^2 \times m^6})^{1/4} , \end{equation} where $\frac{d_I}{m^3} $ is the number density of IBHs at the end of inflation.  $d_I$ is smaller than, but not much smaller than, $1$.  It is as large as it can be consistent with the fact that the black holes do not immediately merge and indeed behave like a dilute black hole gas. 

All of these results are consistent with our demonstration that the HST models can explain the observed crossover between matter and radiation in our universe.  Thus a single model, based on the coarse grained behavior of black holes, can explain inflation, CMB fluctuations, baryogenesis and dark matter.   The two parts of the argument that need much more investigation both involve calculation of merger rates of black holes in a black hole gas in the non-linear regime.  We do not see how to get reliable answers without extensive numerical simulation.  However, the phenomenologically required value of the ratio between PBH and IBH number densities at the reheating transition where most of the IBHs decay is so small, that it is unclear whether even a state of the art N-body simulation could calculate it accurately. We'll explore the implications of this in the conclusions.

The most important question that these models face is whether they are compatible with bounds on black hole decay in the early universe.  In this paper we've imposed only the constraint that the PBHs are stable until they dominate the universe at the observed crossover energy density.  Observation requires that they be stable for more or less the current lifetime of the universe.  This is a mass of order $10^{20} - 10^{21}$ in Planck units and the few decades in mass above that bound are currently free from observational constraints.  The few decades below that bound are strongly constrained, in the sense that $< 0.1\%$ of the dark matter could have been made of such PBHs.   In this paper we've shown that in HST models one can plausibly accomodate PBHs that dominate the universe at the observed energy density of matter radiation equality.  N-body simulations will be required to determine whether those PBHs can combine to form larger ones satisfying all observational bounds. 

\section{Conclusions}

Although the HST cosmological models are based on finite models of quantum gravity compatible with unitarity and "gravitational locality"\footnote{This is an awkward phrase for models that localize subsets of variables inside causal diamonds of finite area/entropy, but do not satisfy the axioms of local quantum field theory.}, actual cosmological results depend, for the most part, only on very general coarse grained properties of those models.  As in all such "hydrodynamical" analyses, there are certain undetermined parameters like susceptibilities, that can take on a variety of values in different models.  At the moment it appears to us the both the slow roll metric $a(t)$ , and the numerical coefficient of $\epsilon^2$ in our prediction for the tensor/scalar ratio $r$
must be included in these model dependent features of our analysis.  We have only been able to impose the weak bound $\epsilon > (10 - {\rm ln}\ (\epsilon)^{-1})$ on the slow roll metric, from the requirement that the universe expand rapidly enough that IBHs do not recombine immediately to form a horizon filling black hole for all time.  

Another disturbing feature of our analysis comes from the equation
\begin{equation} a n_R =  10^{-28} e_R. \end{equation}  In the context of the HST model this equation tells us that only a fraction of order $10^{-23}$ of the IBHs need to combine into PBHs in order to account for the observed universe.  Since each PBH consists of roughly $a m \sim 10^5 a/\epsilon$ IBHs, we need the model to predict only a low probability for the required combination to occur.   The essential lesson of quantum mechanics is that theoretical physicists cannot predict the future.   In a classical analysis of black hole mergers in the period between $(m\epsilon)^{3/2}$ and $(am)^3$ Planck times after the end of inflation, the initial velocity distribution of the IBHs will be important.   This distribution is determined by the quantum dynamics of the underlying model.  It probably varies from model to model.   Even if we have picked out a unique model, we're asking that there be about a $10^{-16} - 10^{-17}$ probability that $IBH \rightarrow PBH$ mergers occur.  Unless the models uniformly predict this probability to be overwhelmingly higher or lower we could not say that this was a "fine tuning issue".   Quantum mechanics tells us that theoretical physicists cannot predict the future with absolute accuracy.   Here we are learning that the value of the dark matter density at the beginning of the matter dominated era is likely to be a quantum mechanical accident, even given a precisely specified mathematical model.
It's also likely that there are many models of the early universe, of equal mathematical consistency, and this is one of the measurements we have to do in order to find out which of them is correct.  This is unfortunate, because there may not be many other tests that pick out a particular model.

\vfill\eject
\vskip.3in
\begin{center}
{\bf Acknowledgments }\\
The work of T. Banks is partially supported by the U.S. Dept. of Energy under grant DE-SC0010008. The work of W.Fischler is supported by the National Science Foundation under Grant Number PHY-1914679.  
\end{center}

\section{Appendix - The Mathematical HST Models}
  
  This paper was based on hydrodynamic reasoning.  We found that most features of cosmology could be explained by arguments involving only entropies of large subsystems of the universe, and general principles of quantum statistical mechanics.  In addition, we imposed the three basic principles of HST: time evolution along individual timelike trajectories in a space-time encoding the global thermodynamics of the underlying quantum system, the quantum relativity principle constraining mutually accessible quantum information in overlapping causal diamonds, and the principle that  objects localized in the bulk of a causal diamond are "dual to" constrained states of the underlying system on the boundary of the diamond, with a relation between the quantum number called energy and the number of constraints.  Energy is not an exactly conserved quantity, but it becomes conserved in the limit of infinite diamonds, where the number of constraints goes to infinity.  Implicit in the definition of localized object is the CEP, which tells us the total number of states in the diamond.
  
The principle that local excitations in the bulk of a diamond are constrained states of a system living on its boundary suggests the form of the Hamiltonian for a theory of quantum gravity.  The two pieces of evidence for this principle from black hole physics are the Schwarzschild de Sitter entropy formula and the huge increase in entropy when a small object is dropped on an existing black hole. In both cases the increase is of order the square root of the entropy of the larger system.  We see no other explanation for these formulae than the idea that two objects that are separated from each other in space live in a Hilbert space larger than the tensor product of the individual system's Hilbert spaces, with a large number of frozen q-dits.  The process of absorption of a localized object into a horizon is the unfreezing of those q-dits and thermalization of the entire system.  

Consistency of this picture requires that, in the state with frozen q-dits, the two other subsystems, local object and horizon, become decoupled.  We know of one class of system where this kind of dynamics is automatic.  If we construct $N \times N$ matrix Hamiltonians, in which each matrix element is an operator in an independent copy of the same q-dit Hilbert space of fixed dimension, then the total entropy of the system 
scales like $N^2$.  The constraint that off diagonal matrix elements between a small subsystem and the rest, vanish on a subspace, freezes of order $N$ degrees of freedom.  Furthermore, it's been argued\cite{hpss} that these systems are fast scramblers, mimicking the dynamics of black holes.  We can arrange that the natural time scale of the dynamics is the horizon radius, by dividing a Hamiltonian obeying 't Hooft scaling by $N$.  

In $4$ space time dimensions, the HST formalism is a special case in which the we take the matrices to be products of an $N\times (N + 1)$ rectangular matrix and its Hermitian conjugate.  The matrix elements are operators in a q-dit Hilbert space, which can always be thought of as the fundamental representation of $SU(p|q) $ with $p + q = d$, and the matrix elements are fermionic generators of the super-algebra.  The Hamiltonian is the trace of an even polynomial in the fermions, with normalization \begin{equation} H = \frac{1}{N} Tr (\frac{\psi^{\dagger\ i}\psi_j }{N} \ldots   \frac{\psi^{\dagger\ k}\psi_l }{N}) C_{ij\ldots kl} + H_{out}. \end{equation}  The coefficients $C$ are order 1 and both the dimension of the q-dit space and rank of the polynomial are kept fixed as $N$ gets large.  The entropy of the Hilbert space is identified with the two dimensional maximal area on the boundary of a causal diamond whose area scales like $N^2$.  In cosmological models, the past tip of the diamond lies on a finite Big Bang Singularity, and the future proper time in the maximal diamond is unbounded.  In an asymptotically dS universe, $N$ asymptotes to a maximal value at infinite proper time.

The full Hilbert space of the model is the one described above, with $N = N_{max}$, which is proportional to the dS radius in Planck units.  $H(N)$ is interpreted as the time dependent Hamiltonian that propagates from the time slice $N - 1$ to the time slice $N$ (in Planck units) in the Milne like coordinates that stay inside the diamond.

Now consider a state where the off diagonal q-dit operators, $\psi_{aB}$ between a block of size $n \ll N_{max}$ and the complementary $N_{max} - n$ block, vanish.  This subspace has an entropy deficit proportion to $N_{max}$ and the probability of a random state having an order one projection on this subspace is $e^{- c n N_{max}}$.   The two blocks behave like non-interacting systems in the constrained subspace, because of the single trace structure of the Hamiltonian.   The large $N$ scaling of the Hamiltonian, tells us that the time scale for equilibration is $N_{max} {\rm ln}\ N_{max}$ in accordance with the behavior of quasi-normal modes on the dS horizon\footnote{With a few small changes of phrase everthing that we are saying here has analogs for a small object dropping into a negative specific heat black hole, with $N_{max}$ replaced by the Schwarzschild radius of the final equilibrated hole.} and the argument of Sekino and Susskind\cite{hpss} that this sort of matrix model is a fast scrambler.  

The paragraph above is also a derivation of (the scaling law for) the Gibbons-Hawking temperature of dS space, {\it if we make the assumption that $n$ is proportional to the energy of the localized object in dS space.}  In order to make this compatible with our quantum Hamiltonian, we have to add a term  $H_c = P_c H(n) P_c , $ where $H(n)$ is the matrix Hamiltonian above, which has energy differences of order $1/n$.  $P_c$ is the projector on the constrained subspace where the $N_{max} \times N_{max}$ matrices are block diagonal.  We have proposed that the radical difference in time scales between these two pieces of the Hamiltonian should be interpreted in terms of redshift.  The Hamiltonian is propagating us in the proper time of a specific timelike geodesic, along Milne like time slices that remain inside the diamond.   States that pass close to the geodesic have energies independent of the diamond radius, while states localized near the boundary of the diamond have very low energy. The quasi-normal mode analysis suggests energy differences for near boundary states should be of order $N_{max}$.  Note that the expectation value of $H(n)$ in generic states is $n$ and energy differences are $o(1/n)$.  Thus $H(n)$ is describing localized objects that can pass close to a detector that lives on the timelike geodesic in the center of the diamond.  

Note that on time scales of order $N_{max}$ the $P_c$ is not a constant of the motion. In a time of order the scrambling time, the projection of the actual state of the system on the constrained subspace has a norm of order $e^{- c N_{max}}$ and the Hamiltonian $H_c$ becomes irrelevant.  

Now let's apply this formalism to cosmology.  We start at the very earliest time with a very small Hilbert space with $N = 1$ .  That space has dimension $d$.  In our fantasies about how string theory compactifications arise in the HST formalism, this space and the fundamental superalgebra of which it is an irreducible representation, encode information about the compact dimensions of space. In the space time picture this earliest time is the Big Bang hypersurface.  In the quantum theory it is non-singular but the hydrodynamic approximation to it, encoded in the FRW cosmology, is not applicable because the entropy is small.   This is the "in" Hilbert space, which is a tensor factor in the full Hilbert space of the system, which is the irreducible representation of $N_{max}(N_{max}+1)$ copies of the fundamental algebra, whose fermionic generators anti-commute with each other.  Every Planck time, we increase the value of $N$ by one, adding a number of generators proportional to $N$ to the algebra.  This increases the dimension of the "in" space and decreases that of the "out" space.  This rule matches the increase in the area of causal diamonds each time the proper time increases by $1$ Planck unit.  If the spatial geometry is flat, the spatial volume inside the causal diamond at time $N$ after the Big Bang is $\propto N^3$, while the entropy is of order $N^2$ and the expectation value of the Hamiltonian is of order $N$.   
We then find entropy and energy densities \begin{equation} \sigma \sim 1/N \sim \sqrt{\rho} . \end{equation} These are the scaling laws and Friedmann equation for the flat FRW cosmology with
equation of state $\rho = p$.  That spacetime has a conformal Killing symmetry in which we rescale time and space by $(\lambda, \lambda^{-3})$ .  This is implemented in the quantum theory by insisting that for large $N$ the Hamiltonian approach that of a $1 + 1$ dimensional CFT, with $c \sim N^2$, living on an interval of length $L$, and with a 
UV cutoff $M$ such that $L M$ is fixed.  The entropy of the CFT then scales like $N^2 LM \propto N^2 {\rm ln}\ d $ and consistency of the field theory approximation requires ${\rm ln}\ d \gg 1$, which we interpret as the claim that the compact spatial dimensions are large in $4 + K$ dimensional Planck units, ${\rm ln}\ d  \sim R_{KK}^K $.  

To describe other geodesics in the same space-time, begin with a tetrahedral lattice of points on the Big Bang hypersurface.  We associate one time-like geodesic with each lattice point.  At time $N$ during the $p = \rho$ era we define the overlap Hilbert space between the geodesics at two lattice points to be the Hilbert space along either geodesic at time $N - s({\bf x, y})$, where $s({\bf x, y})$ is the minimal number of lattice steps between the two points.   The two systems have the same sequence of time dependent Hamiltonians (by definition of a homogeneous model) so if they begin in the same state the density matrices on this tensor factor will be the same.  Notice that during this era we do not have to specify anything about the Hamiltonian that acts on the "out" factor of the Hilbert space.  

In this formalism, each rectangular matrix $\psi_i^J$ is thought of as transforming in the tensor product of the $[N]$ and $[N + 1]$ dimensional representation of $SU(2)$ and is a cutoff (fuzzification\cite{tbhk} ) of the chiral spinor bundle over the two sphere.  We quantize the matrix elements in a way that preserves $SU(2)$.  Products of chiral and anti-chiral spinors are differential forms and a single trace Hamiltonian is the fuzzy analog of the integral of a product of forms over the sphere.  

In order to describe a period of inflation, we need a period of proper time during which the dimension of the "in" Hilbert space does not change.  Let's call that size $N = m_+$.  During this period the actual state of the system can sometimes fit into a constrained Hilbert space in the sense that its overlap with the constrained subspace will be $o(1)$ even though it is a space of much lower dimension.  Thus there will be some average entropy $S = m^2$ with fluctuations
$\delta S/S \sim m^{-1}$.  $m$ defines a time scale $\sim m$ in Planck units, {\it the inflationary Hubble time.}  $m$ will be less than $m_+$, but not too far from it.  

There will also be fluctuations in angular momentum.  These should reproduce the Kerr Black hole entropy formula, which means that they are peaked around $L = 0$ with fluctuations of order $\delta L/S \sim  1/m$.  We cannot calculate the ratio of the size of angular momentum fluctuations to entropy fluctuations without more detailed information about the microscopic Hamiltonian\footnote{In unpublished work with Daniel Park, we showed that the Hilbert space had angular momenta as large as $N^3$ but that the vast bulk of the states had angular momenta $\leq N^2$.  The distribution is peaked around zero but is too flat to reproduce the Kerr black hole entropy formula.  The Hamiltonian must be chosen to reproduce the correct distribution of angular momenta, but this seems like a fairly weak constraint.}.  This ratio would appear to be something like the ratio of two susceptibilities in a condensed matter system, which cannot be calculated using only hydrodynamics.
The usual rules of statistical mechanics tell us that non-Gaussian fluctuations will be suppressed by higher powers of $S^{-1/2}$. 

Now we have to explain the slow roll era in which, from the space-time point of view, the horizon expands to include everything we see today.  In HST, this means simply that in conformal FRW coordinates we are at the exact midpoint between the Big Bang singularity and the singularity in conformal time that corresponds to infinite proper time along our chosen geodesic in asymptotically dS space.  The space-like surfaces of the causal coordinates in our diamond up to this time all lie below this line and intersect the past boundary of the diamond on spheres in the FRW past.  Now let us impose the condition that other timelike geodesics underwent the same period of inflation.  When we see them, we are looking out at a particular angular position in the sky, and for consistency with the description along the other geodesic, we see an isolated quantum system with a given entropy.   This is only possible if the current entropy in our diamond is much larger, so that we can describe isolated systems by saying that the degrees of freedom that give rise to interactions between our detector and that system are frozen.   Here we are using the language of the matrix model.  In space time terms this means that the horizon has expanded.   

Note that the entropy in a causal diamond whose future tip is at $\eta_0/2$ is {\it not} given by the area of the horizon at $\eta_0 /2$.   That area is the full entropy of dS space.
If $n_{bh}$ is the number density of black holes of mass $m$ on the FRW slice at $\eta_0 / 2$ (the end of inflation) then we must have 
\begin{equation}  a^{3} (\eta_0 / 2) (\eta_0 / 2)^{3} n_{bh} m < a(\eta_0 / 2) (\eta_0 /2) . \end{equation}  This is the condition that the Schwarzschild radius corresponding to the black hole energy is less that the horizon radius.   From the point of view of the matrix model, it is the condition that the number of constrained q-dits be smaller than the number of unconstrained horizon q-dits.  Note that for $(a(\eta_0/2) \eta_0/2) \sim m$, which is what we expect for a period of slow roll inflation, where the expansion rate changes only slowly, the bound gives
\begin{equation} n_{bh} < m^{-3} .\end{equation} 
This is the equation we have used in the text.

Our other important constraint on slow roll comes from the requirement that, from the matrix model point of view, the number of q-dits in the "in" factor of the Hilbert space increases sufficiently rapidly, that the IBHs can remain isolated quantum systems.  Here one is talking about {\it the number of IBHs in the causal diamonds with future tips below the line $\eta_0/ 2$}.  This number is small, because those diamonds do not have access to most of the degrees of freedom that are accessible to a detector near conformal time $\eta_0$ and are interpreted by a theorist explaining the data in that detector as "q-dits living on the FRW slice at $\eta_0 / 2$".  The total number of q-dits accessible in a causal diamond with future tip at $\eta$ is $\propto (a(\eta ) \eta)^2$ .
Assume we want to have $o(1)$ isolated subsystems with about $m^2$ q-dits, we must maintain of order $m a(\eta) \eta $ frozen q-dits for a time of order $a(\eta) \eta {\rm ln}\ (m a(\eta) \eta) $ .   The time scale for unfreezing a q-dit is $a(\eta) \eta$, so if the rate of increase of the number of q-dits is faster than 
$$\frac{1}{{\rm ln}\ (m a(\eta) \eta)} $$ 
then the $m^2$ q-dit system will remain independent throughout the slow roll evolution.  This leads to the modest lower bound on the slow roll parameter $\epsilon$ described in the text. For the value of $m = 10^5 \epsilon^{-1}$ needed to fit CMB data, this leads to
\begin{equation} \epsilon > c (10 - {\rm ln}\ (\epsilon))^{-1} . \end{equation} We are not able to determine the numerical constant $c$ in this equation, and it would appear to depend on the precise choice of fast scrambling Hamiltonian in  our model.   This bound is compatible with the phenomenological upper bound on $\epsilon$ obtained from the fit to CMB data,within our ability to calculate.  If the undetermined constants that determine the tensor to scalar ratio $r$, and $c$ above, are indeed model dependent, then the combination of {\it a priori} theoretical constraints, and current data, are compatible if all the unknown constants have values of order $1$.  

Note that there are no constraints on $\epsilon$ from unitarity and locality.  As theorists we can choose to allow the Hamiltonian $H_{in} (t)$ to expand the number of q-dits on which it acts, in any way that we please.  If we want to model a system that makes a transition from inflation to a dilute black hole gas, then we must obey the constraints discussed above.

The only other theoretical constraint we have on our choice of $H_{in} (t)$ and $H_{out} (t)$ is the quantum principle of relativity.  We have already used this twice in our discussion.  We showed that it was compatible with any choice of $H_{out}$ during the inflationary era.  We also used it to explain why, within HST, an inflationary era is always followed by a dilute black hole gas era, {\it if the horizon is allowed to expand rapidly enough for the IBHs to remain as isolated quantum systems}.  During the discrete black hole gas era, since we are describing individual black holes in a completely coarse grained manner, it would seem that the only constraints come from requiring that a black hole entering the past boundary of some trajectory's causal diamond, had to be present as a constraint on the Hilbert space describing nearby causal diamonds.   Obviously the details of this constraint depend on assumptions about the initial velocity distribution of the black holes, a function of all the black hole velocities that we have not had occasion to discuss up until this point.  It is also clear that the question of whether the number density and spectrum of PBHs produced by the model is compatible with observation, the main focus of this paper, depends on this function.  Thus, determining how to calculate this distribution from the matrix models is a central question for the HST cosmological models.  It is a question that we do not know how to address at the current time, and the answer to it is likely to depend on specific details of the matrix model.  Thus, along with the detailed prediction for the slow roll metric $a(t)$, and the numerical coefficients in the tensor scalar ratio, and non-Gaussian fluctuations, the question may eventually just involve fitting models to data, without prospect of further tests for those models.  It is also certain that even given a particular model, the prediction for the number of PBHs of each mass will be a quantum mechanical probability distribution.   Thus, since we have only one universe to study observationally, we may never know whether the model predicts the correct history of the universe.  In this type of model, the point of transition between matter and radiation domination, which determines much of the later fate of the universe, might be the ultimate Schrodinger cat.

Finally we want to outline the argument for approximate de Sitter invariance of CMB fluctuations.  The Friedmann equation and $p =\rho$ equation of state of our model of the early universe follow from the CEP and general large $N$ scaling properties of matrix models. The flatness of the universe requires us to build in another feature namely that for large $N$ the model approaches a $1 + 1$ dimensional CFT.  This is not a fine tuning requirement because random bilinear fermion models approach the free $1 + 1$ dimensional Dirac Hamiltonian\cite{kw}.  Any such model will have an approximate $SL(2,R)$ dynamical symmetry, with the Hamiltonian one of the generators of the group.  If $N$ goes to infinity, the symmetry becomes exact.  

In our inflation model, the fluctuations in the CMB are described by a large number of 
independent systems with this approximate symmetry, and $N = 10^5 / \epsilon$.  

These systems appear as black holes (IBHs) penetrating the past boundary of the maximal conformal diamond of the asymptotically dS universe.   The Hamiltonian describing the dynamics on that boundary is invariant under (fuzzy) area preserving 
maps of the two sphere, a group with many $SU(2)$ subgroups.  The IBHs are 
independent subsystems of the degrees of freedom, which remain approximately 
decoupled until they begin to recombine to form PBHs.  Thus, we are free to identify 
them with small $U(1)$ symmetric caps of the sphere, placed in an almost $SU(2)$ 
invariant manner\footnote{We can make the distribution invariant under a large 
dihedral subgroup of $SU(2)$, the symmetry of a triangularly latticized anti-prism, with 
a number of points equal to the total number of IBHs.}.  Each cap has a small circular 
annulus around 
This uses the map between matrix elements and area elements on the sphere.  The generators of $SU(2)$ combine with
the generators of the local $SL(2,R)$ group, organized as 
\begin{equation} J_{04} = \sum_{\Omega (i)} L_0 (\Omega (i)) , \end{equation}
\begin{equation} {\bf J}_{\pm i} = \sum_{\Omega (i)} {\bf \Omega (i)}  L_{\pm} (\Omega (i)) . \end{equation}
to form an approximate $SO(1,4)$, under which the distribution of fluctuations is approximately invariant.  Localization on the sphere and the consequent emergence of a particular $SU(2)$ subgroup of the group of area preserving maps is what ultimately gives rise to evanescent local physics in the bulk of our universe.  

It is the gradual interaction of the IBHs with each other, which gives rise to the PBHs through merger, to the CMB radiation, the excess of baryons over anti-baryons, and the later mergers of PBHS that give rise to galaxies and stars and planets. This local cosmological history is evanescent.   Along any geodesic of the hydrodynamic FRW spacetime, it will all fade into the cosmological horizon in a time of order $10^{61}$ Planck units, which is $\sim 10^{10}$ earthly years.  By following the accelerated trajectory of our own local group of galaxies, a small remnant of this cosmic history remains visible until the local group collapses into a black hole.  In the HST model, the universe is eternal\footnote{It's possible to construct a meta-physical, mathematical model based on the same HST principles\cite{tbwf??}, which resolves "fine tuning" problems by environmental selection, and eliminates the philosophical "Boltzmann Brain" objection to eternal dS space.  As far as we can tell, this model has no other testable consequences.} but localized objects are "poor players, who strut and fret their hours upon the stage, and then are heard no more." \cite{shakespeare}


\begin{thebibliography}{99}
\bibitem{pbh} B.~Carr and F.~ Kuhnel, 
``Primordial Black Holes as Dark Matter: Recent Developments,"
[arXiv:2006.02838 [astro-ph.CO]] and references therein.
\bibitem{nz} Zeldovich, Ya. B.. Novikov, I. D., Astron. Zh. 43, 758.
\bibitem{hpss} P.~Hayden and J.~Preskill,
``Black holes as mirrors: Quantum information in random subsystems,''
JHEP \textbf{09}, 120 (2007)
doi:10.1088/1126-6708/2007/09/120
[arXiv:0708.4025 [hep-th]]; Y.~Sekino and L.~Susskind,
``Fast Scramblers,''
JHEP \textbf{10}, 065 (2008)
doi:10.1088/1126-6708/2008/10/065
[arXiv:0808.2096 [hep-th]].
\bibitem{ted}  T.~Jacobson,
``Thermodynamics of space-time: The Einstein equation of state,''
Phys. Rev. Lett. \textbf{75}, 1260-1263 (1995)
doi:10.1103/PhysRevLett.75.1260
[arXiv:gr-qc/9504004 [gr-qc]].

\bibitem{tbrant}  T.~Banks,
``On the Limits of Effective Quantum Field Theory: Eternal Inflation, Landscapes, and Other Mythical Beasts,''
[arXiv:1910.12817 [hep-th]].
\bibitem{tbwfbaryo} T.~Banks and W.~Fischler,
``CP Violation and Baryogenesis in the Presence of Black Holes,''
[arXiv:1505.00472 [hep-th]].
\bibitem{kw} V.~Kaplunovsky and M.~Weinstein,
``SPACE-TIME: ARENA OR ILLUSION?,''
Phys. Rev. D \textbf{31}, 1879 (1985)
doi:10.1103/PhysRevD.31.1879
\bibitem{tbhk} T.~Banks and J.~Kehayias,
``Fuzzy Geometry via the Spinor Bundle, with Applications to Holographic Space-time and Matrix Theory,''
Phys. Rev. D \textbf{84}, 086008 (2011)
doi:10.1103/PhysRevD.84.086008
[arXiv:1106.1179 [hep-th]].
\bibitem{tbwf??}
T.~Banks and W.~Fischler, 
``Holographic Inflation Revised", 
[arXiv:1501.01686 [hep-th]]; 
T.~Banks and W.~Fischler, 
``Holographic Space-time, Newton's Law and the Dynamics of Black Holes,"
[arXiv:1606.01267 [hep-th]];
T.~Banks and W.~Fischler, T.J.~Torres,and C. L.~Wainwright,
``Holographic Fluctuations from Unitary de Sitter Invariant Field Theory,"
[arXiv:1306.3999 [hep-th]];
T.~Banks and W.~Fischler,
``Holographic Theories of Inflation and Fluctuations"
[arXiv:1111.4948 [hep-th]];
 T.~Banks and W.~Fischler,
  ``The holographic approach to cosmology,''
  [arXiv:0412097 [hep-th]];
  T.~Banks, W.~Fischler, W. and L.~Mannelli, L, 
  ``Microscopic quantum mechanics of the p = rho universe,"
  PhysRevD.71.123514
  [arXiv:0408076 [hep-th]];
T.~Banks and W.~Fischler,
  ``Holographic cosmology,''
  [arXiv:0405200 [hep-th]];
  %%CITATION = HEP-TH/0405200
  T.~Banks and W.~Fischler,
  ``Holographic cosmology 3.0,''
  Phys.\ Scripta T {\bf 117}, 56 (2005)
  [arXiv:0310288[hep-th]].
  \bibitem{shakespeare} William Shakespeare, ``Macbeth", Act V, Scene 5.
  
  
 

  
  \end{thebibliography}
\end{document}